\documentclass{Interspeech}

\interspeechcameraready

\title{A Self-Training Approach for Whisper to Enhance Long Dysarthric Speech Recognition}

\author[]{Shiyao}{Wang}
\author[]{Jiaming}{Zhou}
\author[]{Shiwan}{Zhao}
\author[affiliation={*}]{Yong}{Qin}

\affiliation[nocounter]{College of Computer Science}{Nankai University}{China}
\email{wangshiyao@mail.nankai.edu.cn}

\usepackage{comment}

\begin{document}

\maketitle

\begin{abstract}
    
    Dysarthric speech recognition (DSR) enhances the accessibility of smart devices for dysarthric speakers with limited mobility. Previously, DSR research was constrained by the fact that existing datasets typically consisted of isolated words, command phrases, and a limited number of sentences spoken by a few individuals. This constrained research to command-interaction systems and speaker adaptation. The Speech Accessibility Project (SAP) changed this by releasing a large and diverse English dysarthric dataset, leading to the SAP Challenge to build speaker- and text-independent DSR systems. We enhanced the Whisper model's performance on long dysarthric speech via a novel self-training method. This method increased training data and adapted the model to handle potentially incomplete speech segments encountered during inference. Our system achieved second place in both Word Error Rate and Semantic Score in the SAP Challenge.
\end{abstract}
\renewcommand{\thefootnote}{}
\footnotetext{* Corresponding author.}

\section{Introduction}
Dysarthria, a speech disorder stemming from neurological conditions like Parkinson's disease and cerebral palsy, affects vocal control and leads to speech impediments such as stuttering, abnormal pauses, altered prosody, and mispronunciation. These characteristics pose significant challenges for automatic speech recognition (ASR) systems, which are primarily trained on typical speech \cite{asrcanotdsr}. Although modern ASR systems \cite{whisper,googleusm} demonstrate robust performance on typical speech due to abundant training data, dysarthric speech recognition (DSR) suffers from data scarcity. This limitation restricts DSR research to word-level \cite{uaspeech} and command-word level \cite{easycall} tasks. Widely used English dysarthric datasets like UASpeech \cite{uaspeech} and Torgo \cite{Torgo} lack the speaker and text diversity necessary for building robust speaker-independent (SI) and text-independent (TI) DSR systems. Consequently, previous DSR research focuses on data augmentation \cite{dda2022,dda2024} and developing personalized models \cite{sd1,2023dsrsurvey,pbdsr,pblrdwws}. To address the data scarcity issue and stimulate more advanced research, Hasegawa-Johnson et al. \cite{sap} launched the Speech Accessibility Project (SAP) to collect a large and diverse dataset of dysarthric English speech and subsequently organized the SAP Challenge.

\begin{figure}[t]
\setlength{\abovecaptionskip}{0.1cm} 
\centerline{\includegraphics[width=8cm]{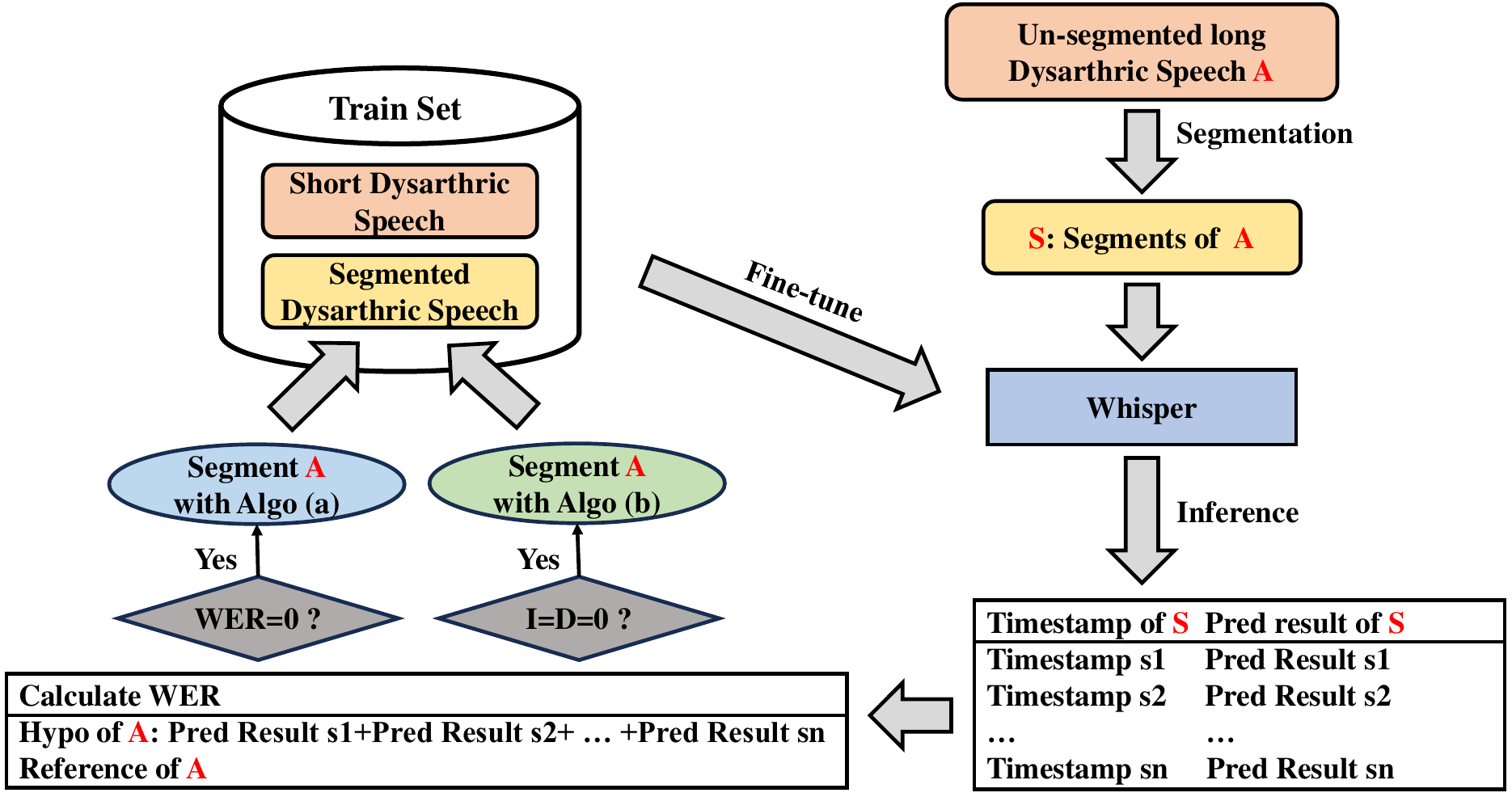}}
\caption{The overview of our self-training approach to segment long dysarthric speech (ST-SLDS). `Hyo': `hypothesis'. Algo (a) and Algo (b) see Fig \ref{fig:segmentation algo}.}
\label{fig:overview}
\vspace{-23pt}
\end{figure}

\renewcommand{\thefootnote}{\arabic{footnote}} 
As participants in the SAP Challenge, we developed SI and TI DSR models using the SAP 2024-04-30 (SAP0430\_processed) dataset\footnote{Sample durations range from 0.039 seconds to 121.02 seconds.}. Building on the findings of Zheng et al. \cite{sap1005interspeech2024} and Singh et al. \cite{sap1005icassp2025}, who demonstrated the effectiveness of fine-tuning models like wav2vec 2.0 \cite{wav2vec2} and Whisper \cite{whisper} on previous SAP data (with Whisper showing superior performance), we prioritized base model selection. We compared Whisper against other advanced ASR technologies not yet evaluated on the SAP data, including HuBERT \cite{HuBERT}, and pre-trained models from ESPnet \cite{espnet} and Wenet \cite{wenet}. Due to Challenge time constraints and GPU utilization efficiency, we trained on short speech segments (see Section \ref{sec:expup} for specific training speech duration limits). Furthermore, because Whisper's audio encoder truncates long spectrograms to fit its positional encoding (leading to incomplete predictions)\cite{m2rwhisper}, we also evaluated model performance with an even segmentation (E-S) strategy during inference (detailed in Section \ref{sec:infer set}). Ultimately, the fine-tuned Whisper large-v3, with the E-S strategy employed for inference, demonstrated superior performance (Table \ref{tab:asr compre}). While Whisper \cite{whisper} uses a buffering approach for long-speech transcription, segmenting audio into 30-second chunks and using predicted timestamps to advance the window, WhisperX \cite{whisperx} has shown that Whisper's timestamp predictions are often inaccurate, limiting long-speech recognition. WhisperX combines timestamp-free decoding with Voice Activity Detection (VAD) for improved results. Specifically, WhisperX pre-segments audio using VAD, concatenates VAD segments into approximately 30-second blocks, and then performs transcription. Inspired by WhisperX, we explored various inference strategies based on timestamp-free decoding (Table \ref{tab:inference setting test}).

While various strategies can improve long-speech inference, Whisper's architecture limits fine-tuning to data segments of 30 seconds or less. This constraint introduces several drawbacks: (1) Long-speech data loss: Manual segmentation and transcript matching are possible \cite{trainonly302}, but the labor cost is excessive. (2) Training-inference mismatch: Segmenting long speech during inference can lead to incomplete sentences in the input audio. As Niehues et al. \cite{imcompleteml} demonstrated in machine translation, models trained only on complete data tend to generate complete sentences even from partial inputs. Liu et al. \cite{imcompleteasr} observed similar mismatches in speech recognition. Both studies addressed this by incorporating ``partial sequence" data into training. Following their approach, we attempted to incorporate segmented data to mitigate the training-inference inconsistency. We segmented the unused long dysarthric speech and annotated it using a self-training method. 

\begin{figure}[t]
\setlength{\abovecaptionskip}{0.1cm} 
\centerline{\includegraphics[width=8cm]{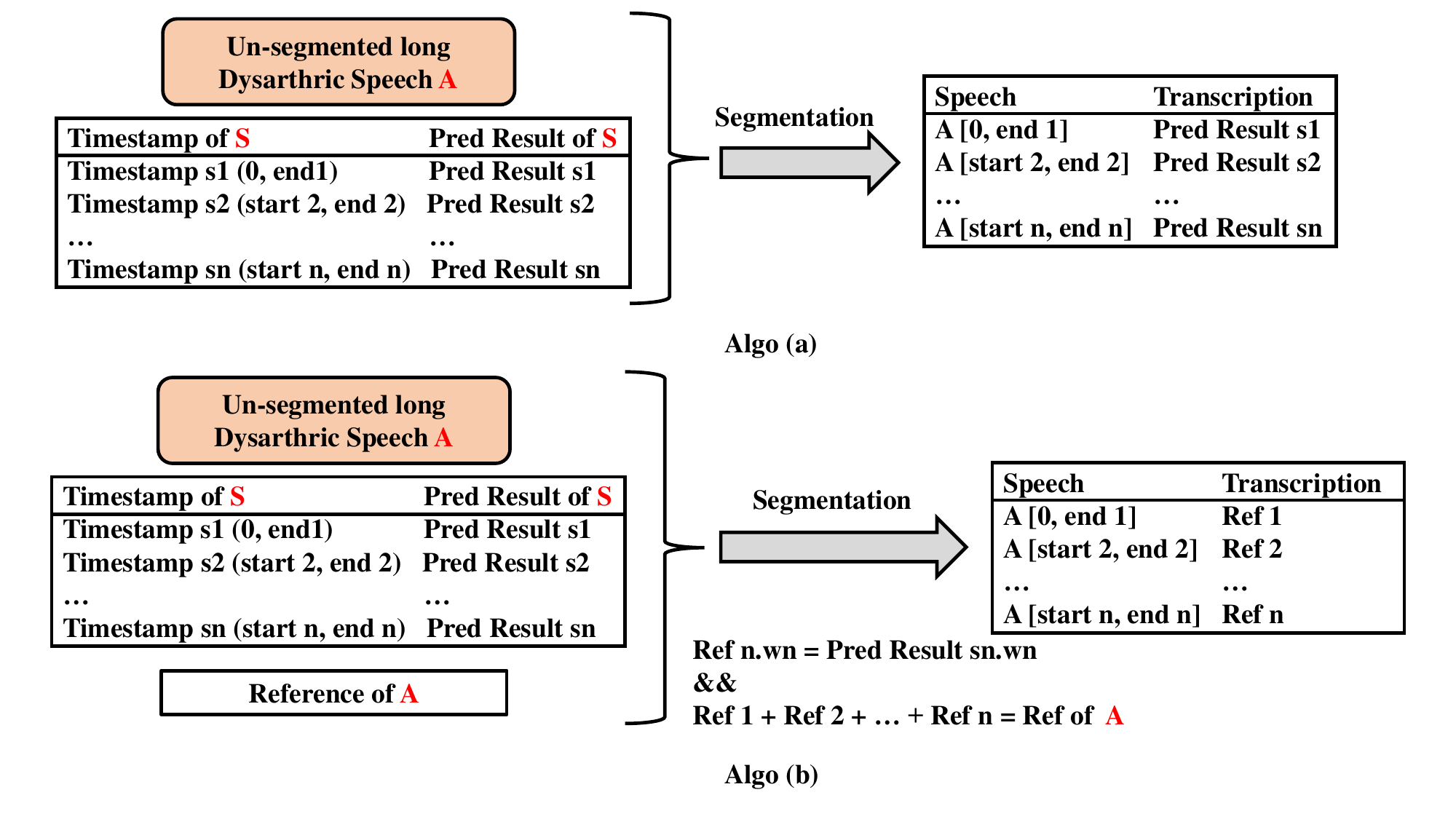}}
\caption{Segmentation algorithm (a) and (b). `wn' represents the number of words. `Ref': `Reference'. `Algo': `Algorithm'.}
\label{fig:segmentation algo}
\vspace{-21pt}
\end{figure}

Self-training, a common semi-supervised learning method \cite{selftrainingfore2easr,selftrainingaccess2019,selftraininglesson,selftraingonlowresource,selftrainingfromstss,madi}, is employed when labeled data is less and unlabeled speech data is abundant. It aims to transform unlabeled data into usable training material, thereby enhancing the model's generalization, robustness, and adaptability. Self-training operates within a teacher-student framework. This framework involves training a teacher model on labeled data, generating pseudo-labels for unlabeled data, and then filtering the unlabeled data based on the confidence of the pseudo-labels \cite{selftraingss} (e.g., frame-level, utterance-level) or through heuristics \cite{selftrainingfore2easr}. The filtered data is then combined with the labeled data to train a student model, which subsequently becomes the new teacher, and the process is repeated iteratively. In the context of improving Whisper's performance on long dysarthric speech, we applied this self-training concept. Specifically, we designed an iterative training process within a teacher-student framework (Figure \ref{fig:overview}). First, we built a teacher model using the short duration speech sample data. Then, we carried out segmentation-based inference on the long audio to generate pseudo-labels. Subsequently, we screened the data based on a heuristic strategy and refined the annotations using designed segmentation algorithms (Figure \ref{fig:segmentation algo}). Finally, we added the segmented data to the training set for a new iteration. 

The main \textbf{contributions} of this work are as follows:
\begin{itemize}
    \item We tested different advanced ASR technologies and various inference settings to build a speaker-independent and text-independent DSR model with strong generalization ability.
    \item We propose a self-training approach specifically designed to enhance Whisper's ability to process long dysarthric speech. This method simultaneously increases the available training data and mitigates the training-inference mismatch, leading to improved performance.
\end{itemize}

\section{Whisper self-training optimization}
\label{sec:self-training}

\setlength{\tabcolsep}{0.7mm}{
\begin{table}[t]
    \setlength{\abovecaptionskip}{0cm} 
    \caption{Performance comparison in WER (\%) among different ASR models. `E-S': even segmentation. `W-' models are different size variants of the Whisper model.}
    \label{tab:asr compre}
    \centering
    \begin{tabular}{l c c c c}
        \toprule
        \textbf{Model} & \textbf{HuBERT} & \textbf{ESPnet} & \textbf{ESPnet} & \textbf{Wenet}\\  
        \textbf{Setting} & \textbf{CTC} & \textbf{Pretrain 1} & \textbf{Pretrain 2} & \textbf{Pretrain}\\
        \midrule
        \textbf{Params} & 94.5 M & 148.9 M & 148.9 M & 131.3 M\\
        \textbf{FT} & 17.7800 & 10.6203 & \textbf{8.8852} & 8.8986\\
        \textbf{FT + E-S} & 18.0746 & 25.3420 & 24.2847 & 8.8700\\
        \toprule
        & \textbf{W-Small} & \textbf{W-Medium} & \textbf{W-Large-v2} & \textbf{W-Large-v3} \\
        
        \midrule
        \textbf{Params} & 241.7 M & 763.6 M & 1543.3 M & 1543.5 M\\
        \textbf{FT} & 16.9325 & 14.3460 & 13.4699 & 13.2869\\
        \textbf{FT + E-S} & 12.3040 & 9.7766 & 8.0043 & \textbf{7.6106}\\
        \bottomrule
    \end{tabular}
  \vspace{-20pt}
\end{table}
}

We propose a self-training approach to segment long dysarthric speech (ST-SLDS, see Figure \ref{fig:overview}). First, we fine-tune the Whisper model on short dysarthric speech data to create a teacher model. In subsequent iterations, the fine-tuning dataset includes segmented data from previous iterations. The teacher model then performs inference on long dysarthric speech. Prior to inference, the long speech is segmented using a segmentation strategy to obtain timestamped segments, which are then individually fed into the teacher model to generate corresponding predictions. To select high-quality data, we use a heuristic screening strategy: (1) We reconstruct a prediction for the original long speech by combining the predictions of its segments, and then calculate its WER against the reference transcription. (2) We select data with either a WER of 0 (WER = 0) or a non-zero WER but zero insertion and deletion errors (I = D = 0). Data with a non-zero WER is re-tested in later iterations. The selected data is then segmented and annotated using the algorithms detailed in Figure \ref{fig:segmentation algo}. Data filtered by the ``WER = 0" condition uses Algorithm (a), segmenting the speech at the predicted timestamps and using the corresponding prediction as the label. Data filtered by the ``I = D = 0" condition uses Algorithm (b), also segmenting at the predicted timestamps but generating labels by extracting the same number of words from the reference transcription, matching the segment's position in the original speech. Finally, the segmented data is added to the training set, and the process is repeated iteratively.

\section{Experiments}
\label{sec:exp}
\subsection{Dataset}
\label{sec:dataset}
The SAP Challenge data comprises the following components: (1) SAP0430\_processed: This dataset has been preprocessed through 16kHz audio resampling and text regularization and is divided into training (290.4 hours, 131,420 samples, 369 speakers) and development (dev) (43.6 hours, 19,275 samples, 55 speakers) partitions. The dev partition was further subdivided by us into a validation set (9,275 samples, 55 speakers) and the 0430Test set (10,000 samples, 54 speakers). Our models are trained using disfluency-removed transcripts. (2) SAP-1130 (released on November 30, 2024): This dataset consists of unprocessed 48kHz audio and unregulated text. We preprocessed the data newly added in the SAP-1130 training partition compared to the SAP0430\_processed training partition. This preprocessed data was then used as the 1130Test set (11 hours, 2,160 samples, 27 speakers). (3) The SAP Challenge evaluates submissions on Test 1 (7,601 samples) and Test 2 (8,043 samples). These evaluation sets are inaccessible to us and exhibit no speaker or text overlap with the data we obtained. 

\begin{figure}[t]
\setlength{\abovecaptionskip}{0.15cm} 
\centerline{\includegraphics[width=7cm]{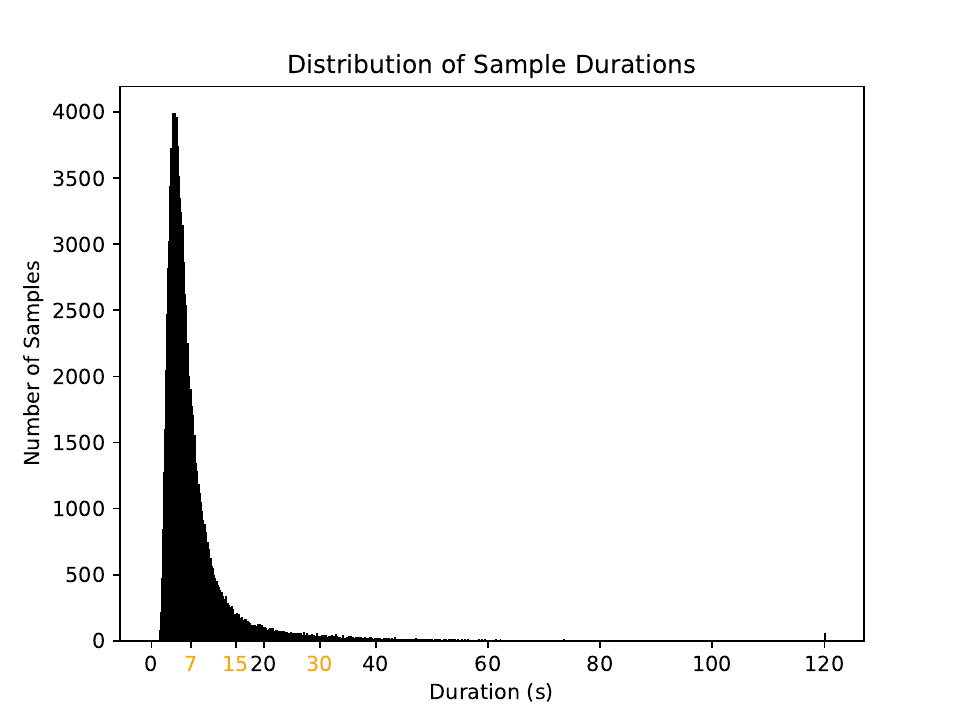}}
\caption{The duration distribution of the training data in SAP0430\_processed.}
\label{fig:duration distribution}
\vspace{-23pt}
\end{figure}
\subsection{Experimental settings} 
\label{sec:expup}
\subsubsection{Model and training settings}
\label{sec:model and train settings}
\noindent\textbf{HuBERT CTC:} We utilized the pre-trained HuBERT Base model\footnote{\url{https://dl.fbaipublicfiles.com/hubert/hubert_base_ls960.pt}} and incorporated a CTC layer to build DSR model. Training was conducted under the Fairseq framework\footnote{\url{https://github.com/facebookresearch/fairseq/blob/main/examples/hubert/README.md}}, based on the hyperparameters outlined in the base\_10h.yaml configuration. We set the learning rate to $10^{-5}$ and the warmup steps to 32,000, halting the training when the validation set loss did not decrease over 10 consecutive epochs.

\noindent\textbf{ESPnet Pretrain:} We fine-tuned two separate models within the ESPnet framework. The first model was pre-trained on the 960-hour LibriSpeech dataset\footnote{\url{https://huggingface.co/asapp/e_branchformer_librispeech}} (ESPnet Pretrain 1), while the second was pre-trained on the 10,000-hour GigaSpeech dataset\footnote{\url{https://huggingface.co/pyf98/gigaspeech_e_branchformer}} (ESPnet Pretrain 2). Both models were fine-tuned for 100 epochs using the train\_asr\_e\_branchformer.yaml configuration.

\noindent\textbf{Wenet Pretrain}: We fine-tuned a model provided by Wenet, which was pre-trained on 10,000-hour GigaSpeech data\footnote{\url{https://github.com/wenet-e2e/wenet/blob/main/docs/pretrained_models.md}}, for 100 epochs using the pre-trained configuration file. 

\noindent\textbf{Whisper} By leveraging the Whisper-Flamingo codebase\footnote{\url{https://github.com/roudimit/whisper-flamingo}}, we fine-tuned several Whisper pre-trained models (small, medium, large-v2, large-v3). All models were fine-tuned using hyperparameters specified in the corresponding configuration files (see config/audio/ in the repository). Given that Whisper \cite{whisper} was pre-trained on original collected text, during which machine-transcribed data (all-upper or all-lower) was filtered out. To align with Whisper's pre-training data characteristics, we evaluated two transcription formats during training: (1) A-U: all uppercase, a common format for ASR model fine-tuning; (2) F-U: the first letter in uppercase and the rest in lowercase.  

\noindent\textbf{Training speech duration limits} Regarding models with fewer parameters compared to Whisper, we limited the maximum speech length to 30 seconds and maximized the batch size on an A100 GPU. The large parameter count of the Whisper large-v3 model restricted the maximum batch size to 2 under the 30-second duration limit \cite{whisperreproduce}, leading to slow training progress and insufficient GPU utilization. By analyzing the duration distribution of the training data in SAP0430\_processed dataset (Figure \ref{fig:duration distribution}), we set a 15-second duration limit for Whisper, which is twice the duration of the highest data concentration.

\setlength{\tabcolsep}{0.5mm}{
\begin{table}[t]
    \setlength{\abovecaptionskip}{0cm} 
    \caption{Performance comparison in WER (\%) and SemScore among different inference settings. A detailed description of the inference settings can be found in Section \ref{sec:expup}. }    
    \label{tab:inference setting test}
    \centering
    \begin{tabular}{c c | c c | c c c | c c}
    \toprule
    \multicolumn{2}{c|}{\textbf{Txn}} & \multicolumn{2}{c|}{\textbf{Segmentation}} & \multicolumn{3}{c|}{\textbf{Decoding}} & \multicolumn{2}{c}{\textbf{Result}}\\
    \textbf{A-U} & \textbf{F-U} & \textbf{E-S} & \textbf{VAD-S} & \textbf{G-S} & \textbf{B-S} & \textbf{Prompt} & \textbf{WER} & \textbf{SemScore}\\
    \midrule
    $\checkmark$ &  & $\checkmark$ & & $\checkmark$ & & & 7.6106 & 92.3187\\
    $\checkmark$ &  & $\checkmark$ & & $\checkmark$ & & 1 & 7.8871 & 92.3996 \\
    $\checkmark$ &  & $\checkmark$ & & & 5 & & 7.4095 & 92.3653 \\
    $\checkmark$ &  & $\checkmark$ & & & 10 & & \textbf{7.3694} & 92.4093\\
    $\checkmark$ &  & $\checkmark$ & & & 10 & 1 & 7.4504 & 92.4863 \\
    $\checkmark$ &  & $\checkmark$ & & & 10 & 2 & 7.5096 & \textbf{92.4873}\\
    $\checkmark$ &  & $\checkmark$ & & & 20 & & 7.3932 & 92.4088\\
    $\checkmark$ &  & $\checkmark$ & & & 30 & & \textbf{7.3685} & 92.4324\\
    \hline
    & $\checkmark$ & $\checkmark$ & & & 10 & & 6.8422 & 93.2347 \\
    & $\checkmark$ & & $\checkmark$ & & 10 & & \textbf{6.3245} & 93.5297 \\
    & $\checkmark$ & & $\checkmark$ & & 10  & 1 & 6.3884 & \textbf{93.5883} \\
    \bottomrule
\end{tabular}
  \vspace{-18pt} 
\end{table}
}
\begin{table*}[t] 
    \setlength{\abovecaptionskip}{0cm} 
    \caption{
    The effect of enhancing Whisper by using our ST-SLDS approach. `1st'-`4th' denote the 1st-4th iterations of the self-training method. `Add dev' means incorporating the dev part of SAP0430\_processed into the training set.
    }
    \label{tab:st-slds} 
    \centering
    \begin{tabular}{l | c | c c | c c | c c | c c}
        \toprule
\textbf{Settings} & \textbf{Train} & \multicolumn{2}{c|}{\textbf{0430Test}} & \multicolumn{2}{c|}{\textbf{1130Test}} & \multicolumn{2}{c|}{\textbf{Test 1 Result}} & \multicolumn{2}{c}{\textbf{Test 2 Result}}\\
& \textbf{Samples} & \textbf{WER} & \textbf{SemScore} & \textbf{WER} & \textbf{SemScore} & \textbf{WER} & \textbf{SemScore} & \textbf{WER} & \textbf{SemScore} \\
        \midrule
        Baseline & 119845 & 6.8422 & 93.2347 & 8.8376 & 89.1004 & 9.2663 & 88.3354 & 11.1954 & 84.9895\\ 
        1st WER=0 & 124556 & 5.8011 & 93.7636 & 7.9411 & 89.9006 & --- & --- & --- & --- \\
        1st WER=0 + I=D=0 & 128094 & 5.7878 & 93.8621 & 7.8882 & 90.0194 & --- & --- & --- & ---\\
        2nd WER=0 & 128974 & 5.6362 & 93.9558 & 7.6172 & 90.3785 & --- & --- & --- & ---\\ 
        2nd WER=0 + I=D=0 & 133988 & 5.5628 & \textbf{94.1113} & 7.5798 & 90.2510 & 9.1918 & 88.3758 & 11.1487 & 85.0233\\
        3rd WER=0 & 130624 & 5.6162 & 93.8716 & 7.6150 & 90.2384 & --- & --- & --- & ---\\
        3rd WER=0 + I=D=0 & 136325 & \textbf{5.5533} & 93.9591 & \textbf{7.4675} & \textbf{90.4589} & 8.7301 & 89.3272 & 10.3923 & 85.9151\\
        4th WER=0 & 131339 & 5.6162 & 93.9340 & 7.6040 & 90.3504 & --- & --- & --- & --- \\
        4th WER=0 + I=D=0 & 137415 & 5.5571 & 93.9712 & 7.5754 & 90.2702 & --- & --- & --- & --- \\
        \hline
        Add dev & 137248 & 3.4635 & 97.9971 & 9.0667 & 89.1355 & --- & --- & --- & --- \\
        1st WER=0 , Add dev & 141959 & 3.1079 & 98.1854 & 8.1283 & 89.7760 & --- & --- & --- & ---\\ 
        1st WER=0 + I=D=0 , Add dev & 145496 &  3.0088 & 98.1669 & 7.7758 & 89.9184 & --- & --- & --- & ---\\ 
        2nd WER=0 , Add dev & 147444 & 2.9392 & 98.1726 & 7.6877 & 90.3077 & --- & --- & --- & ---\\
        2nd WER=0 + I=D=0 , Add dev & 153509 & 2.6170 & 98.4272 & 7.6415 & \textbf{90.3769} & 8.4151 & 89.6910 & 10.3091 & 86.0170\\
        
        3rd WER=0 , Add dev & 150970 &  2.8820 & 98.2746 & 7.6327 & 90.1662 & --- & --- & --- & ---\\
        3rd WER=0 + I=D=0 , Add dev & 158246 & 2.5559 & 98.4155 & \textbf{7.5820} & 90.0180 & \textbf{8.2479} & \textbf{89.7438} & \textbf{10.0294} & \textbf{86.1884}\\ 
        4th WER=0 , Add dev & 152358 & 2.5502 & \textbf{98.4969} & 7.8001 & 89.7829 & 8.5954	& 89.2178 & 10.4831 & 85.6587\\
        4th WER=0 + I=D=0 , Add dev & 160072 & \textbf{2.5493} & 98.4856 & 7.6459 & 90.2306 & 8.5507 & 89.1929 & 10.3536 & 85.7684\\
        \bottomrule
    \end{tabular}
    \vspace{-18pt}
\end{table*}

\subsubsection{Inference strategies}
\label{sec:infer set}

\noindent\textbf{Segmentation strategies:} 
\begin{itemize}
    \item \textbf{Even Segmentation (E-S)}: Long speech is evenly partitioned into segments. The number of segments ($n$) is determined by: 
    \begin{align}
        n&= N // (L_{max}\times16000)+1,
        \label{equation:eq1}
    \end{align}
    where $N$ is the number of audio sampling points and $L_{max}$ is the maximum input duration.
    \item \textbf{VAD Segmentation (VAD-S)}: We utilize Silero VAD\footnote{\url{https://github.com/snakers4/silero-vad}} to generate the timestamps of speech segments. In contrast to WhisperX, we preserve all segments, even those without detected speech. Segmentation begins at timestamp 0, with subsequent points determined by the VAD's detected speech start timestamps. An iterative process ensures the time interval between segmentation points remains close to, but does not exceed $L_{max}$. Given that Silero VAD is not specifically trained for dysarthric speech, its detection may be inaccurate, and in some cases, it may fail to detect any speech segments. If this occurs, we employ the E-S strategy for segmentation.
\end{itemize}

\noindent\textbf{Decoding strategies:} 
\begin{itemize}
    \item Greedy Search (G-S)
    \item Beam Search (B-S): Beam size is a configurable parameter (e.g., 5, 10, 20, 30).
    \item Prompt: For segmented speech, the result of the previous segment is used as a prompt for Whisper's inference of the current segment.
\end{itemize}

\subsubsection{Evaluation metrics}
The SAP Challenge uses two evaluation metrics: Word Error Rate (WER) and Semantic Score (SemScore). WER quantifies the normalized edit distance between the reference transcript and the DSR system's output. SemScore combines BERTScore \cite{bertscorefordsr}, phonetic distance, and natural language inference probability into a single metric. A lower WER and a higher SemScore indicate superior performance. See the Challenge's template repository\footnote{\url{https://github.com/xiuwenz2/SAPC-template}} for detailed metric calculations.

\subsection{Results and discussions}
\label{sec:result}
Table \ref{tab:asr compre} compares the performance of DSR models fine-tuned from different pre-trained models on 0430Test.
In direct inference (without segmentation), the fine-tuned ESPnet Pretrain 2 model achieved the best results. However, Whisper's performance was limited by its encoder's truncation of long spectrograms, resulting in incomplete predictions. To accurately assess Whisper's potential, we implemented a simple E-S strategy during inference. This significantly improved Whisper's performance (p \textless 0.05), with the fine-tuned Whisper large-v3 model achieving the best results using the E-S strategy. Conversely, applying the E-S strategy degraded the performance of other models, except for Wenet Pretrain. Finally, we selected fine-tuned Whisper large-v3 as our base model and adopted the segmentation strategy for long speech during inference.

Table \ref{tab:inference setting test} presents a comparison of the performance of different inference settings on 0430Test.
To isolate the impact of decoding strategies, we first tested them using the A-U fine-tuned model and the simple E-S strategy. Beam search outperformed greedy search, with a beam size of 10 proving optimal. Although increasing the beam size to 30 slightly decreased the WER, local testing indicated that it exceeded the Challenge's 240-minute limit. Implementing the Prompt strategy with a beam size of 10 increased the SemScore but worsened the WER. Connecting the predictions of the previous two segments as a prompt (Prompt = 2) further increased the SemScore slightly, yet the negative impact on the WER was so substantial that we only considered Prompt = 1. Subsequently, we tested improved inference strategies on the model fine-tuned with the F-U setting, which yielded significantly better performance than the A-U fine-tuned model, indicating the advantage of aligning the fine-tuning setting with the pre-training setting. Replacing the E-S strategy with the VAD-S strategy, which provides better segmentation points, further enhanced performance. We also tested inferring only speech segments, similar to WhisperX (The results were not presented in Table \ref{tab:inference setting test}). This approach produced performance inferior to that of the VAD-S strategy. Presumably, since the VAD tool was not adapted to dysarthria, it failed to accurately detect speech segments in dysarthric speech. Therefore, we chose to fine-tune the model based on the F-U setting, set the beam size to 10 during inference, and use the VAD-S strategy. 
We didn't adopt the Prompt strategy as it had slow inference and worsened WER. 

We further improved performance by augmenting the training data using our ST-SLDS method to segment long dysarthric speech and incorporating the SAP0430\_processed dev partition (including 0430Test).
Initially, tests were conducted without the dev partition. As shown in Table \ref{tab:st-slds}, during the first three iterations, the WER for both 0430Test and 1130Test steadily decreased, indicating the effectiveness of the ST-SLDS method. However, in the fourth iteration, this trend changed. The remaining long dysarthric speech was highly challenging, making it difficult to meet our filtering conditions. Consequently, minimal new data was added. Furthermore, due to our method's re-filtering of ``I = D = 0" data in the next iteration, duplicate data emerged in the new data pool. We hypothesize that these duplicates induced overfitting in the model during the fourth iteration, leading to a performance rebound. Therefore, we halted further iterations. Subsequently, we introduced the dev partition data into the training process. Since 0430Test was integrated into the training set, its performance is no longer a reference point. Notably, the performance of 1130Test under this new setting was lower compared to the settings without the dev partition. We attribute this to the dev partition substantially increasing the number of speakers and the overall complexity of the learning task. Given the existing speaker overlap between 1130Test and the original training set, the altered speaker distribution likely led to the decline in model performance. We submitted the models with relatively superior performance under both settings (with and without the dev partition). The results from Test 1 and Test 2 clearly demonstrate the effectiveness of both our strategies for augmenting the training data. Furthermore, submitting the model from the fourth iteration with the dev partition conclusively verified that three iterations represented the optimal choice.

\section{Conclusions}
\label{sec:conclusions}
To improve Whisper's recognition of long dysarthric speech, we propose a self-training method that segments and labels long speech, enabling its use for fine-tuning and mitigating training-inference mismatch. By segmenting the data, we simulate the incomplete speech segments encountered during inference. Ultimately, our constructed speaker-independent and text-independent dysarthric speech recognition system ranked second in both the WER and SemScore of the SAP Challenge.
\section{Acknowledgements}
This work has been supported in part by  NSF China (Grant No.62271270).

\bibliographystyle{IEEEtran}
\bibliography{mybib}
\end{document}